\pdfoutput=1
\documentclass[letterpaper]{article} 
\usepackage{aaai2026}  
\nocopyright
\usepackage{times}  
\usepackage{helvet}  
\usepackage{courier}  
\usepackage[hyphens]{url}  
\usepackage{graphicx} 
\usepackage{amsmath}
\usepackage{natbib}  
\usepackage{caption} 
\usepackage{booktabs}
\usepackage{algorithm}
\usepackage{algorithmic}

\usepackage{newfloat}
\usepackage{listings}

\DeclareCaptionStyle{ruled}{labelfont=normalfont,labelsep=colon,strut=off} 
\floatstyle{ruled}
\newfloat{listing}{tb}{lst}{}
\floatname{listing}{Listing}
\title{Let the Bees Find the Weak Spots: A Path Planning Perspective on Multi-Turn Jailbreak Attacks against LLMs}

\author{Yize Liu\textsuperscript{1}, Yunyun Hou\textsuperscript{1}, Aina Sui\textsuperscript{1}\thanks{Corresponding author.}}
\affiliations{
\textsuperscript{1} School of Computer and Cyberspace Security, Communication University of China, Beijing, China\\

}

\usepackage{bibentry}

\begin{document}

\maketitle

\begin{abstract}
Large Language Models (LLMs) have been widely deployed across various applications, yet their potential security and ethical risks have raised increasing concerns. Existing research employs red teaming evaluations, utilizing multi-turn jailbreaks to identify potential vulnerabilities in LLMs. However, these approaches often lack exploration of successful dialogue trajectories within the attack space, and they tend to overlook the considerable overhead associated with the attack process. To address these limitations, this paper first introduces a theoretical model based on dynamically weighted graph topology, abstracting the multi-turn attack process as a path planning problem. Based on this framework, we propose ABC, an enhanced Artificial Bee Colony algorithm for multi-turn jailbreaks, featuring a collaborative search mechanism with employed, onlooker, and scout bees. This algorithm significantly improves the efficiency of optimal attack path search while substantially reducing the average number of queries required. Empirical evaluations on three open-source and two proprietary language models demonstrate the effectiveness of our approach, achieving attack success rates above 90\% across the board, with a peak of 98\% on GPT-3.5-Turbo, and outperforming existing baselines. Furthermore, it achieves comparable success with only 26 queries on average, significantly reducing red teaming overhead and highlighting its superior efficiency.
\end{abstract}

\section{Introduction}

Large Language Models (LLMs) such as ChatGPT, Claude, and Gemini have exhibited remarkable generalization performance in natural language processing and have been widely adopted across a variety of applications, including dialogue systems and code generation~\cite{achiam2023gpt}. However, as large-scale deployment of these models continues to expand, concerns regarding their security have become increasingly prominent ~\cite{BARMAN2024dark}. Attackers can exploit specific techniques~\cite{yi2024jailbreak,li2024llm} to induce models into generating harmful, discriminatory, or ethically inappropriate content, thereby posing significant risks to public safety. Although researchers have proposed various techniques, such as reinforcement learning and Direct Preference Optimization (DPO)~\cite{rafailov2023direct} to enhance model alignment and safety~\cite{dai2023safe,liu2023training}, existing defenses remain insufficient to fully prevent malicious attacks.

Currently, security evaluation of LLMs primarily relies on red teaming~\cite{ganguli2022red,openai2023redteam}, which identifies model vulnerabilities by simulating adversarial behaviors. Despite the lack of standardized evaluation criteria, two key objectives have gradually become widely recognized in practice: one is to maximize the \textit{Attack Success Rate} (ASR) so as to comprehensively expose security weaknesses across diverse scenarios; the other is to minimize the \textit{average number of queries}, thereby reducing computational resource consumption and enhancing evaluation efficiency. Early red teaming efforts predominantly employed single-turn attacks, which trigger model violations through one-time adversarial prompt construction, including manual generation~\cite{anil2024many}, white-box attacks~\cite{zou2023universal}, and LLM-assisted attacks~\cite{sitawarin2024pal,shah2023scalable,yu2023gptfuzzer}. However, with continuous advancements in model alignment techniques, the success rate of single-turn attacks has significantly declined, limiting their effectiveness in reliably identifying model vulnerabilities.

Recent studies have demonstrated that multi-turn attack methods can effectively overcome many of the limitations faced by single-turn attacks in identifying model vulnerabilities~\cite{chao2025jailbreaking}. By emulating prolonged adversarial interactions with target models in realistic dialogue scenarios, these methods allow for dynamic refinement and adaptation of attack strategies, significantly enhancing their strategic flexibility, targeting precision, and overall effectiveness. A growing body of research on multi-turn jailbreak attacks~\cite{du2025multi,ding2023wolf} adopts strategies that fragment adversarial intent across multiple dialogue turns, progressively uncovering and exploiting security vulnerabilities through iterative interactions. In addition, several studies have integrated specially trained attacker agent models into the multi-turn dialogue process, where they serve as key components for actively steering the evolution of attack strategies. These agents are typically fine-tuned on red-teaming objectives, enabling them to detect exploitable signals in model responses and generate contextually adaptive follow-up prompts, thereby improving the overall efficiency and success rate of the attack chain.

Despite the increased flexibility and success rates demonstrated by multi-turn red teaming in practice, current research still faces two key challenges. \textbf{First}, there is insufficient attention to the overall \textit{attack space}. Existing studies typically perform repeated attack attempts centered on the same query, utilizing proposed methods to generate multiple distinct attack strategy combinations from the available strategy set. We define the set of all possible attack strategy combinations as the attack space, encompassing all strategy sequences accessible to the attacker. Current approaches often lack systematic exploration of the attack space and overlook the potential of mining attack paths from the global structure of strategy combinations. \textbf{Second}, high resource consumption and limited practicality remain significant obstacles. Some studies rely on dedicated attack models for strategy selection, which require substantial computational resources for training and iterative refinement. Others adopt exhaustive methods that repeatedly invoke attack strategies to uncover potential vulnerabilities. Both approaches significantly increase attack costs and restrict the applicability of these methods in resource-constrained environments.

To bridge this gap, this paper introduces a multi-turn jailbreak attack method based on a dynamically weighted graph topology, and proposes an enhanced ABC algorithm that significantly reduces resource consumption while ensuring high ASR. Specifically, we formalize the search strategy within the attack space as a \textit{path planning} problem on a directed weighted graph. In the constructed graph, nodes represent attack strategies, edges denote the combinatorial relationships between strategies, and edge weights correspond to the evaluated effectiveness of the respective strategy sequences. The attack objective is thus transformed into searching for an optimal path on the graph that is both short in length and high in attack success probability.

Building on this, we propose an enhanced ABC algorithm composed of employed, onlooker, and scout bees, which collaboratively exploit localized exploration and global solution tracking to efficiently identify high-success paths within the large attack space. To accommodate the characteristics of multi-turn attacks, we optimize several core components of the algorithm. In particular, the fitness function is discretized into five levels, and for responses in the intermediate tiers, we extract and utilize informative cues to refine the guidance of subsequent attack strategies. This design enhances both the granularity of path planning and the overall ASR. Extensive experiments were conducted on the JailbreakBench benchmark~\cite{chao2024jailbreakbench}, involving three open-source and two closed-source models. The results demonstrate that our method achieves significantly higher ASR than existing baselines on most models, while markedly reducing the average number of queries. This not only enhances attack effectiveness but also markedly lowers the cost of red teaming evaluations.

Our contributions are as follows:
\begin{itemize}
    \item We introduce a theoretical model based on dynamically weighted graph topology to characterize the multi-turn attack space, formalizing the multi-turn attack problem as a path planning task.
    \item We propose an enhanced Artificial Bee Colony (ABC) algorithm that enables dynamic and efficient search within the attack space, with a fitness function specifically optimized for multi-turn attack scenarios.
    \item We conduct extensive experiments on the latest benchmark datasets involving five large language models, demonstrating that our method significantly outperforms existing baselines in both ASR and average number of queries, thereby validating the superiority of the proposed approach.
\end{itemize}

\section{Related Work}
Research on the safety of large language models (LLMs) primarily comprises defenses and red teaming evaluation. Training-time defenses employ preference alignment and safety fine-tuning to improve refusal consistency. Inference-time defenses rely on system prompts and policy templates, as well as input and output filtering or rewriting, to enforce constraints during interaction. Red-teaming complements defenses by systematically uncovering failure modes in realistic conversational settings and quantifying the safety–utility trade-off. Against this backdrop, jailbreak attacks~\cite{wei2023jailbroken} are a key adversarial technique for revealing alignment vulnerabilities. By crafting covert and targeted inputs, attackers can bypass built-in safeguards and induce the model to produce content that may violate ethical norms, laws, or platform policies. Based on interaction mode and turn count, jailbreak attacks are generally classified into single-turn and multi-turn variants.

\textbf{Single-turn jailbreak attacks.} Single-turn jailbreak attacks construct adversarial prompts in a one-shot manner to elicit unsafe outputs from LLMs. These methods generally fall into two categories. White-box approaches leverage gradient-based optimization to manipulate model inputs or representations, as in GCG~\cite{zou2023universal}, which applies greedy coordinate descent to generate effective adversarial prompts. Black-box methods rely on semantic manipulations or algorithmic heuristics without gradient access. Examples include AutoDAN~\cite{liu2024autodan}, SelfCipher~\cite{yuan2024cipher}, and PAP~\cite{zeng2024johnny}, which adopt diverse optimization strategies. PRS~\cite{andriushchenko2025jailbreaking} further performs random search over suffixes appended to benign prompts to maximize the target log-likelihood of unsafe completions. Despite their simplicity, single-turn attacks have become less effective due to improved alignment techniques, limiting their utility for evaluating models under stronger defenses.

\textbf{Multi-turn jailbreak attacks.} Compared with single-turn attacks, multi-turn jailbreak attacks emulate realistic adversarial interactions by engaging the target model in iterative dialogues, gradually inducing it to produce boundary-violating outputs. PAIR~\cite{chao2025jailbreaking} uses dialogue-driven iterative refinement to achieve semantic jailbreaks under black-box access, whereas CoA~\cite{sun2024multiturncontextjailbreakattack} couples a seed generator with toxicity escalation to preconstruct and adapt multi-turn attack chains in response to model feedback. Crescendo~\cite{russinovich2025crescendo} introduces a progressive multi-turn dialogue approach that begins with benign queries and gradually leverages model outputs from previous turns to intensify the attack. GOAT~\cite{pavlova2024automatedredteaminggoat} employs an attacker agent that adaptively selects attack strategies based on the target model’s responses during the multi-round interaction. GALA~\cite{chen2025strategizegloballyadaptlocally} proposes a novel dual-dimensional learning mechanism, training the attacker agent using both global strategic planning and local prompt-level refinement.

\textbf{Path Planning and Swarm Intelligence.} Path planning formulates decision making as finding shortest or constraint-satisfying paths on weighted graphs. Classical explicit methods such as Dijkstra typically operate under static edge costs and admissible heuristics. When costs depend on interaction history, explicit search often suffers from frontier blow-up and fragile heuristics. Swarm-intelligence approaches such as PSO and ABC maintain diversity through parallel neighborhood exploration, probabilistic selection, and restart mechanisms, and are widely used in combinatorial routing and navigation, offering a practical toolkit for long-horizon problems with expensive evaluations.

Despite gains in ASR and realism, prior multi-turn studies largely optimize strategy choice within individual dialogues and rarely model history-dependent costs at a global level. Other lines rely on attacker-agent training or extensive query-based search, which increases evaluation cost and constrains scalability under realistic budgets. These gaps motivate adopting a planning-style attack search that achieves higher ASR while controlling attack cost.

\section{Methodology}

In this section, we provide a comprehensive description of the proposed dynamically weighted graph topology model and the enhanced \textit{Artificial Bee Colony} (ABC) algorithm. Our approach abstracts the search for strategy combinations in multi-turn attacks as a path planning problem on a dynamically weighted graph topology and utilizes the enhanced ABC algorithm to intelligently explore and optimize attack paths.

\subsection{Attack Space Modeling}
To systematically characterize the multi-turn jailbreak process, we model the attack space as a directed weighted topology and, on this basis, construct a layered state graph to capture path-dependent memory.

\textbf{Graph and paths.}  We model the multi-turn attack space as a  directed complete graph topology  $G=(V,E)$. The vertex comprises the initial query node together with a set of attack-strategy nodes, and edges denote the sequential combinational relationships among these strategies. Attack strategies are sourced from prior work~\cite{lin2024achillesheel,jiang2024wildteaming} and community repositories, and are organized into several classes according to content characteristics. A path of length $T$ is denoted $p=(v_0\to v_1\to\cdots\to v_T)$, where $v_0$ denotes the initial harmful query.

\textbf{Path score.} In this model, an edge weight is defined as the harmfulness score induced on the target model by a successor edge under a given prefix history. Let the path prefix be $\operatorname{pref}(p,T) = (v_0 \to v_1 \to \cdots \to v_{T-1})$, and let the terminal edge be $e_T = (v_{T-1} \to v_T)$. Then, the resulting score of the entire path is defined as
\begin{equation}
S_T = w(\operatorname{pref}(p,T), e_T),
\label{eq:score}
\end{equation}
where \(w\) denotes the target model’s harmfulness-scoring function under the current combination of strategies. When \(S_T\) reaches the harm threshold \(a\), the attack is considered successful. The objective is to find a successful attack path of minimal length.

\textbf{Layered state graph.} To better represent that the same edge may take different weights under different historical prefixes, we extend the directed complete graph into a layered state graph so that path memory is encoded as a static topology in the state space. We introduce a progress mapping $P(\operatorname{pref}(p,t)) \in \{0,\dots,a\}$, which compresses any historical prefix into a finite progress level. Based on this, we construct layered state graph as
\begin{equation}
\hat{\mathcal{G}}=(\hat{V},\hat{E}), \quad
\hat{V}=\{(v,s)\mid v\in V,\ s\in\{0,\dots,a\}\},
\label{eq:layered_graph_short}
\end{equation}
were $(v,s)$ denotes the state of being at strategy $v$ with prefix progress level $s$. If there exists a prefix $\operatorname{pref}(p,t)$ in the original graph with $P(\operatorname{pref}(p,t))=s$, and after extending by node $u$ the new progress becomes
\begin{equation}
s'=\min\{a,\; P(\operatorname{pref}(p,t),\, u)\}.
\label{eq:state_transition_no_concat}
\end{equation}
then an edge $(v,s)\to(u,s')$ exists in $\hat{\mathcal{G}}$.

\textbf{Search objective.} We cast the goal of multi-turn jailbreaks as a path-planning problem over the layered state graph. Let the start state be $(v_0,0)$, where $v_0$ is the initial harmful query, and let the goal set be $\{(v,a)\mid v\in V\}$. Within this finite, static space formed by strategy nodes and progress levels, the task is to reach any success state $(v,a)$ in the fewest steps, equivalently, to solve a unit-weight shortest-path problem on $\hat{\mathcal{G}}$:
\begin{equation}
\min_{p \subseteq \hat{\mathcal{G}}} |p|
\quad \text{s.t.} \quad
(v_0,0) \xrightarrow{p} (v,a).
\label{eq:search_objective}
\end{equation}
This reduction converts the original history dependence into static transitions across a finite set of levels while remaining strictly aligned with the attack’s success criterion.

\subsection{Theoretical Properties and Rationale}
Building on the graph modeling and problem analysis, we employ an improved ABC-based swarm optimization. Compared with traditional meta-heuristics, ABC is better aligned with this task: it treats multiple candidate prefixes as parallel units, reuses high-quality prefix fragments during local edits, and employs probabilistic selection to mitigate single-instance misjudgments. Under an attack-dominated cost model, ABC reaches the success level in fewer rounds with lower implementation overhead. Meanwhile, when the theoretical graph topology is manageable, explicit shortest-path search (BFS) performs well. But in our setting, edge weights are highly prefix-dependent, and explicit expansion causes the state space and frontier to grow rapidly. By contrast, ABC on an implicit layered space avoids enumerating long histories and, through parallel probing, raises the per-round probability of effective progress, thus achieving higher efficiency and greater stability under limited evaluation budgets. We next analyze ABC’s typical properties in the multi-turn jailbreak setting as follows.

Let \(m\) denote the population size, \(a\) the success-threshold level, \(s_0\) the initial progress level, \(\eta\) the minimum effective per-step level increment, \(q\) a uniform lower bound on the per-round progress probability, \(t\) the number of effective parallel trials per round, and \(C_{\mathrm{eval}}\) the per-evaluation model time cost.

\textbf{Time complexity.} We bound the expected total time by
\begin{equation}
\mathrm{E}[T]
= O\!\left(
C_{\mathrm{eval}}\, t \sum_{\ell = s_0}^{a-1} \frac{1}{\eta_{\ell}\, q_{\ell}}
\right).
\end{equation}
Here, the summation term estimates the expected number of progress rounds from \(s_0\) to \(a\), and each round incurs \(t\) evaluations at a per-evaluation cost of \(C_{\mathrm{eval}}\).

\textbf{Space complexity.} The memory usage scales as
\begin{equation}
\mathrm{Space}_{\mathrm{ABC}}=O\!\left(m\,\frac{a-s_0}{\eta}\right).
\end{equation}
For comparison, explicit shortest-path search requires \(O(na)\) space with \(n = |V|\), whereas ABC grows only linearly with the population size \(m\) and the number of levels.

\textbf{Convergence.} We obtain the following convergence bound:
\begin{equation}
\mathrm{E}[\tau_a] \;\le\; \sum_{\ell = s_0}^{a-1} \frac{1}{\eta_{\ell}\, q_{\ell}} .
\end{equation}
Here, \(\tau_a\) denotes the hitting time to the success level \(a\). The layered state space is finite and the success level is absorbing. With reuse of high-quality prefix fragments and a restart mechanism, these conditions yield the above upper bound on the expected hitting time.

\subsection{ABC Algorithm for Multi-Turn Attacks}

The Artificial Bee Colony (ABC) algorithm is a representative swarm intelligence optimization method and a widely adopted approach for solving path planning problems. In this study, we employ ABC to efficiently search for candidate attack paths given an initial query, aiming to maximize success rates while minimizing query overhead and achieving an effective balance between performance and cost. The ABC algorithm consists of three primary phases: the employed bee phase, the onlooker bee phase, and the scout bee phase. The overall procedure is illustrated in Algorithm~\ref{alg:abc}.
The population size determines the number of employed and onlooker bees, where each individual in the population corresponds to a candidate path and is associated with either an employed or an onlooker bee.

\begin{algorithm}[t]
\caption{Artificial Bee Colony for Multi-Turn Attack}
\label{alg:abc}
\begin{algorithmic}[1]
\REQUIRE Initial population $P = \{r_1, r_2, \ldots, r_l\}$; max iterations $N$; update threshold $\gamma$; selection size $m$
\ENSURE Best attack path $r^*$
\FORALL{$r \in P$}
    \STATE $\text{UpdateCount}(r) \gets 0$
    \STATE $\text{Fitness}(r) \gets \text{CalculateFitness}(r)$
\ENDFOR
\FOR{$t = 1$ to $N$}
    \STATE $S \gets \text{Selection}(P, m)$ \hfill // Extract high-quality paths from $P$
    \STATE $P \gets (P \setminus S) \cup \{ \text{Mutation}(r) \mid r \in S \}$
    \FORALL{$r \in P$ newly generated by Mutation}
        \STATE $\text{Fitness}(r) \gets \text{CalculateFitness}(r)$
    \ENDFOR
    \FORALL{$r \in P$}
        \IF{$\text{UpdateCount}(r) \ge \gamma$ \AND $\text{Fitness}(r) < \text{MaxFitness}(P)$}
            \STATE $r \gets \text{GenerateNewRoad()}$
            \STATE $\text{Fitness}(r) \gets \text{CalculateFitness}(r)$; $\text{UpdateCount}(r) \gets 0$\hfill // Reinitialize stagnated path
        \ENDIF
    \ENDFOR
    \IF{$\exists r \in P$ such that $\text{Fitness}(r) = \text{Target}$}
        \STATE \textbf{break}
    \ENDIF
\ENDFOR
\RETURN $\text{Best}(P)$
\end{algorithmic}
\end{algorithm}

\begin{table*}[t]
\centering
\small
\renewcommand{\arraystretch}{1.2} 
\setlength{\tabcolsep}{8pt} 
\begin{tabular}{lccccc}
\toprule
\textbf{Method} & \textbf{LLaMA-2-7B} & \textbf{LLaMA-3.1-8B} & \textbf{LLaMA-3.1-70B} & \textbf{GPT-3.5} & \textbf{GPT-4} \\
\midrule
GCG        & 3\%     & /     & /     & 47\%     & 4\%      \\
PAIR       & 0\%     & /     & /     & 71\%     & 34\%     \\
PRS        & \underline{90\%} & / & /     & \underline{93\%} & 78\% \\
Crescendo  & 72\%    & 84.5\% & 77\% & 80.4\%   & 70.9\%   \\
GOAT       & 85.3\%  & \textbf{96.5\%}    & 91\%     & 91.6\%   & \underline{87.9\%}   \\
GALA       & /       & 87\%  & \underline{92\%}   & 91\%     & /        \\
\textbf{ABC (Ours)} & \textbf{96\%} & \underline{94\%} & \textbf{94\%} & \textbf{98\%} & \textbf{90\%} \\
\bottomrule
\end{tabular}
\caption{
Comparison of Attack Success Rate (ASR) across seven methods on five target models. 
The best and second-best results are highlighted in bold and underlined, respectively. 
The performance data for GCG, PAIR, and PRS were sourced from the official JailbreakBench leaderboard, 
while the results for Crescendo and GOAT were computed based on the number of successful queries out of 10 multi-turn attempts.
}
\label{tab:asr}
\end{table*}

In the employed bee phase, the algorithm first evaluates the fitness of each candidate attack path in the population, which measures the effectiveness of the path in the initial stage of the attack. This phase aims to construct the initial search space and lay the foundation for subsequent path optimization. The fitness function scores each path based on the harmfulness of the target model’s response to the final prompt in the path, thereby reflecting the attack potential of the current path. In the onlooker bee phase, the algorithm selects a subset of high-quality paths from the population according to their fitness scores, using a predefined ratio and probabilistic selection mechanism to form the candidate pool. Each candidate path is then subjected to a mutation process to further expand the search space and improve the algorithm’s ability to escape local optima. The mutation operations include random replacement, where an attack strategy within the path is substituted to explore alternative combinations, and random insertion, where a new strategy node is added to enhance the path’s expressive capacity and increase its attack depth. In the scout bee phase, the algorithm monitors the activity and contribution of each individual in the population. If a path has not been updated over multiple iterations and its fitness remains suboptimal, it is considered to be stuck in a stagnation state. In such cases, the scout bee mechanism is triggered to restructure the local search space. Specifically, the stagnant path is replaced with a newly generated random path, thereby enhancing search diversity and helping the algorithm avoid local optima. The components of the algorithm are described in detail as follows.

\textbf{Population Initialization.} During the entire algorithm execution, the population size is kept constant. Initially, we construct a population composed of multiple paths, where the attack strategies in each path are uniformly and randomly selected from a predefined strategy set. To accommodate the need for search space expansion in subsequent stages, while also minimizing the number of queries in multi-turn attacks, the initial path length is restricted to no more than three strategy nodes. The generation process of the population and paths can be formalized as:

\begin{equation}
P = \{ r_1, r_2, \ldots, r_l \}, \quad r_i = (t_{i1}, t_{i2}, \ldots, t_{ik})
\end{equation}

\noindent where each $t_{ij}$ is independently drawn from a uniform distribution over the predefined strategy set $T$, i.e., 
$t_{ij} \sim \mathcal{U}(T)$, and the initial path length is constrained such that $|r| \leq 3$.

\textbf{Selection.} The selection module employs a tournament selection mechanism. Specifically, the algorithm randomly selects a subset of paths from the entire population to form a candidate set. Within this candidate set, the path with the highest fitness score is chosen as the winner of the current tournament. This process is repeated multiple times to progressively filter out higher-performing paths, thereby constructing the final selection set. The selection set can be represented by the formula:

\begin{equation}
S = \left\{ \arg\max_{r \in R_j} \text{Fitness}(r) \right\}_{j=1}^n
\label{eq:selection}
\end{equation}

\noindent where $R_j$ is the candidate subset for the $j$-th tournament, consisting of $m$ paths randomly sampled from the population, and $n$ tournaments are conducted in total.

\textbf{Mutation.} Mutation is a crucial module for maintaining population diversity and discovering superior paths. For each selected path, the algorithm performs one of two mutation operations based on a probability coefficient. The first operation is random replacement, which involves selecting a new attack strategy at a random position in the path to replace the original strategy, thereby exploring new possible strategy combinations. The second operation is random insertion, where a new attack strategy node is appended at the end of the path, ensuring that the total path length does not exceed a predefined maximum limit. Both operations allow the newly added strategies to duplicate existing ones within the path, expanding the search space and increasing the algorithm’s chances of finding the optimal path. The mutation operation \( m(r) \) can be represented as follows:

\begin{equation}
m(r) =
\begin{cases}
(t_1, \ldots, t_{p-1}, t', t_{p+1}, \ldots, t_k), & \beta \\
(t_1, \ldots, t_k, t'), & 1 - \beta
\end{cases}
\end{equation}

\noindent where $t'$ is sampled from a uniform distribution over the predefined strategy set $T$, i.e., $t' \sim \mathcal{U}(T)$, and the augmented path in the second case satisfies $|t_1, ..., t_k, t'| \leq L_{\max}$, where $L_{\max}$ is the maximum allowed path length.

\textbf{Fitness Evaluation.} Previous studies~\cite{russinovich2025crescendo} have shown that embedding harmful queries in batches across multiple dialogue turns can significantly improve ASR. Inspired by this approach, we design a targeted fitness function to finely evaluate the harmfulness of the model's response at each turn. Specifically, the fitness function categorizes model responses into five levels: level 1 indicates a completely safe response with no harmful content, while level 5 represents a fully harmful output generated by the model. Levels 2 through 4 correspond to responses containing partial harmful content, from which the attacker can extract useful information to construct more targeted adversarial prompts for the next turn, progressively advancing the attack toward full success. To implement this mechanism, we design a dedicated prompt template that instructs the model to extract information from its response that is useful for subsequent attack prompts, simulating a semantic extraction process in an uncensored environment. The extracted key content is then combined with the initial query to form the input for the next turn, enabling a gradual and iterative progression of the multi-turn attack.

\begin{table*}[t]
\centering
\small
\renewcommand{\arraystretch}{1.2}  
\setlength{\tabcolsep}{8pt}       
\begin{tabular}{lccccc}
\toprule
\textbf{Method} & \textbf{LLaMA-2-7B} & \textbf{LLaMA-3.1-8B} & \textbf{LLaMA-3.1-70B} & \textbf{GPT-3.5} & \textbf{GPT-4} \\
\midrule
Crescendo       & 50    & 50    & 50    & 50    & 50    \\
GOAT            & 50    & 50    & 50    & 50    & 50    \\
\textbf{ABC (Ours)} & \textbf{27.9} & \textbf{27.7} & \textbf{27.5} & \textbf{10.3} & \textbf{36.6} \\
\bottomrule
\end{tabular}
\caption{
Average number of queries for multi-turn attack methods across five target models. 
Lower values indicate higher attack efficiency. The query numbers listed for baseline methods correspond exactly to the ASR results in Table~\ref{tab:asr}. Note that the 50-query budget for GOAT and Crescendo is strictly derived from the evaluation protocol specified in their original papers: for each harmful seed, the attacker independently interacts with the target model $k=10$ times, with 5 turns per interaction, and an attack is considered successful if at least one of the 10 dialogues produces an unsafe output.
}
\label{tab:queries}
\end{table*}


\section{Experiment}

\subsection{Experimental Setup}

\textbf{Dataset.} We adopt JailbreakBench~\cite{chao2024jailbreakbench} as the primary benchmark to evaluate the effectiveness of our proposed method. This dataset comprises 100 diverse samples of harmful behaviors. Among them, 55\% are originally collected instances, while the remaining samples are carefully curated from existing benchmark datasets such as AdvBench~\cite{zou2023universal} and HarmBench~\cite{mazeika2024harmbenchstandardizedevaluationframework}. All behavior categories in the dataset are aligned with OpenAI's safety policies, spanning a wide range of high-risk domains, including but not limited to phishing, privacy violations, hate speech, and illegal activities. This makes JailbreakBench a more advanced and comprehensive benchmark for assessing the robustness and safety of language models under adversarial scenarios.

\textbf{Scoring Setup.} We employ StrongREJECT~\cite{souly2024strongrejectjailbreaks} as the evaluator to assess the harmfulness of model outputs. Compared to existing evaluation tools, StrongREJECT offers a more rigorous assessment by jointly considering two critical dimensions: the target model’s willingness to refuse harmful queries and its tendency to provide actionable or harmful content. In our experiments, we use an open-source version of StrongREJECT fine-tuned on Gemma 2B, which retains the high accuracy of rubric-based evaluation while significantly reducing computational overhead. This makes it well-suited for our resource-constrained scenario, enabling efficient and reliable evaluation of adversarial outputs.

\textbf{Evaluation Metrics.} We adopt the Attack Success Rate (ASR) as the primary evaluation metric, defined as the proportion of successfully attacked samples relative to the total number of test cases. Additionally, to assess the resource efficiency of different algorithms within the same category, we report the average number of queries required per successful attack as a secondary metric.

\textbf{Attacker LLM.} The attacker language model is used to assist in generating adversarial prompts, and its primary requirement is low sensitivity to safety constraints when faced with potentially harmful instructions. In this work, we employ a non-aligned ``helpful-only'' model Gemma-9B-uncensored as the adversarial prompt generator. Trained on a mixture of both benign and harmful data, the model is designed to prioritize instruction-following without regard for safety constraints, as it has not been subjected to any safety fine-tuning or filtering. Importantly, no explicit red-teaming samples or jailbreak-specific data are included during its pretraining or fine-tuning. This ensures that its adversarial capabilities in our experiments arise purely from the structure and semantics of the input prompts, rather than from memorization or prior exposure to attack-specific examples.

\textbf{Target LLMs.} We conducted evaluations on both open-source and closed-source models. For open-source models, we selected LLaMA family members, including LLaMA 2 7b ~\cite{touvron2023llama2openfoundation}, LLaMA 3.1 8b, and LLaMA 3.1 70b, to assess the effectiveness of our method across different parameter scales. For closed-source models, we selected GPT-4-Turbo~\cite{achiam2023gpt} and GPT-3.5-Turbo, both of which feature advanced safety mechanisms. Evaluating our method on these models offers a more rigorous validation of its practical effectiveness.

\textbf{Algorithm Hyperparameters.} To strike an effective balance between attack efficiency and query cost, we configured two distinct sets of hyperparameters tailored to different model groups. For smaller-scale models, including LLaMA 2 7b, LLaMA 3.1 8b, and GPT-3.5-Turbo, we capped both the number of iterations and maximum path length at 3, with a population size of 5. For the larger-scale model group, such as LLaMA 3.1 70b and GPT-4-Turbo, due to their increased complexity and higher attack difficulty, we appropriately increased the number of iterations and path length to 5, and expanded the population size to 10, thereby enhancing the search space and the algorithm’s exploration capability to improve attack effectiveness.

\begin{figure*}[t]
\centering
\includegraphics[width=\linewidth]{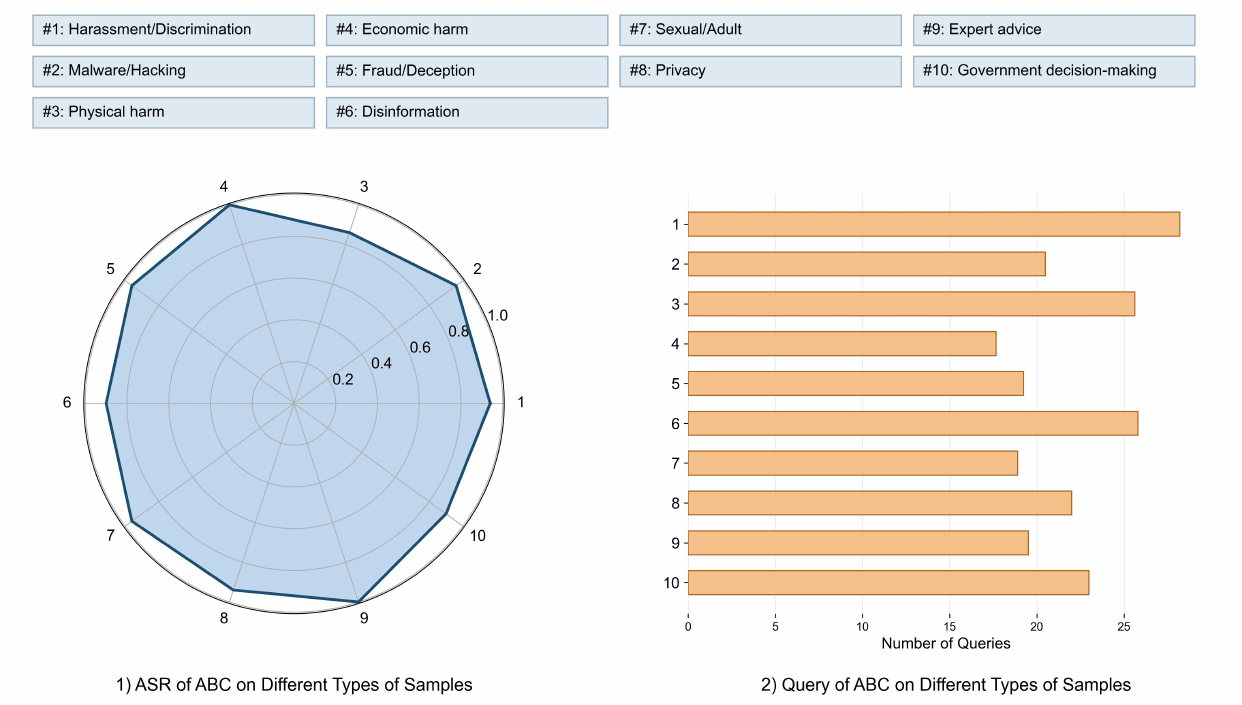}
\caption{Attack performance of ABC across different harm categories. Left: Attack Success Rate (ASR); Right: Average Number of Queries. The radar plot and horizontal bar chart respectively summarize ABC’s effectiveness and efficiency across 10 types of harmful behaviors defined in JailbreakBench.}
\label{fig:radar}
\end{figure*}

\textbf{Baselines.} We compare the proposed ABC algorithm with several jailbreak attack methods, including single-turn attacks such as GCG~\cite{zou2023universal} and PRS~\cite{andriushchenko2025jailbreaking}, as well as multi-turn attacks like PAIR~\cite{chao2025jailbreaking}, Crescendo~\cite{russinovich2025crescendo}, GOAT~\cite{pavlova2024automatedredteaminggoat}, and GALA~\cite{chen2025strategizegloballyadaptlocally}.Among these, Crescendo and GOAT perform repeated multi-turn attacks to conceal harmful queries, while GALA leverages dual-dimensional learning with trained agents to assist in the attack process.

\subsection{Results}

We conduct a systematic comparison between ABC and several baseline approaches from three key dimensions: effectiveness, efficiency, and generalizability. The evaluation focuses on three specific metrics: ASR, average number of queries, and performance across diverse attack scenarios. These metrics align with the core objectives of jailbreak attacks: ensuring attack success, minimizing testing costs, and enhancing generalization. The ASR reflects the method's ability to bypass existing safety mechanisms, while the average number of queries captures the resource cost associated with executing the attack.  Additionally, fine-grained evaluations across different attack types help reveal the method’s stability and adaptability in varied scenarios, offering important insights into its real-world applicability. In the following sections, we analyze the results from these three perspectives.

\subsubsection{Attack Success Rate}

We compare our method with seven baseline approaches across five target models, with the results summarized in Table~\ref{tab:asr}. As shown in the table, in the single-turn attack setting, PRS achieves a success rate of 93\% on GPT-3.5-Turbo, standing out among baseline methods. In the multi-turn scenario, GOAT and GALA perform relatively well, with both methods attaining an average ASR close to 90\% across the five target models. In contrast, ABC demonstrates even stronger performance, consistently outperforming all baseline approaches on nearly every target model. This highlights its superior capability in red-teaming evaluations. Notably, ABC achieves over 90\% success on all tested models, reaching as high as 96\% on LLaMA 2 7b and 98\% on GPT-3.5-Turbo. These results indicate that our method not only offers high attack effectiveness but also exhibits strong robustness, making it highly valuable for practical applications such as red-teaming, safety assessment, and the optimization of defense strategies.

\subsubsection{Average Number of Queries}

To demonstrate that ABC not only significantly improves ASR but also reduces evaluation overhead in multi-turn dialogue settings, we conduct a statistical analysis of the average number of queries used in our experiments. Note that the 50-query budget reported for GOAT and Crescendo is strictly derived from the evaluation protocol in their original papers, where each harmful seed is queried $k=10$ times with 5 turns per interaction. The results are presented in Table~\ref{tab:queries}. As shown in the comparison results, ABC consistently achieves lower average number of queries than the strong baselines Crescendo and GOAT across all five target models. On GPT-3.5-Turbo, for example, it requires only 10 queries on average to complete a successful jailbreak, compared to 50 for the baselines. Even on the more complex GPT-4-Turbo, ABC maintains a low query cost, significantly reducing the overall burden of red-teaming while effectively balancing ASR and query efficiency.

Moreover, other multi-turn attack approaches, such as GALA, rely on training dedicated attacker models to generate adversarial prompts. While this can improve ASR, it introduces considerable resource and time costs due to model fine-tuning. In contrast, our ABC method efficiently searches for attack paths through optimization alone, without any attacker model training or auxiliary learning modules. This substantially reduces both implementation barriers and system overhead, making ABC more lightweight and easier to deploy in practical red teaming. These advantages further highlight its superior efficiency and usability over existing baselines.

\subsubsection{Diverse Scenario Analysis}

To further evaluate the generalization capability of our proposed method in diverse attack scenarios, we conduct a fine-grained analysis of ABC using the JailbreakBench dataset. This benchmark categorizes all test samples into ten representative types of harmful behavior, enabling a systematic assessment of the method's adaptability and robustness across different attack categories.

We aggregate the ASR of ABC across all target models according to these ten scenario categories and visualize the results. As shown in Figure~\ref{fig:radar}, ABC consistently maintains an ASR above 85\% across all categories, and even exceeds 95\% in particularly challenging cases such as malware and hacking. These results demonstrate that ABC not only achieves strong overall attack performance but also exhibits consistent and reliable effectiveness across a wide range of adversarial intents, highlighting its generality and practical utility.

Further analysis shows that the top five categories with the highest ASR also correspond to the five categories with the lowest average number of queries. This phenomenon is likely influenced by both the characteristics of the strategies in the initial strategy pool and the security mechanisms employed by the target models. In these scenarios, the tested models exhibit weaker defenses against jailbreak attacks, indicating an urgent need for strengthening and optimizing protective measures. Notably, this result also reveals an interesting relationship between ASR and average number of queries overhead: as the ASR increases, the required number of queries actually decreases. This observation differs from the common perception that these two metrics are typically at odds. Although further experimental validation is needed, achieving higher attack effectiveness while reducing resource consumption remains the core objective of red team testing. The performance of the ABC method strongly aligns with this goal, highlighting its significant practical value in security assessment.

\section{Limitation and Future Work}

Despite the strong empirical performance of our enhanced Artificial Bee Colony (ABC) algorithm across multiple large language models, several limitations remain. \textbf{First}, the initial population in our current implementation is generated purely at random. While this ensures diversity to some extent, it may lead to a lack of high-quality candidate paths in the early stages, potentially hindering convergence efficiency and overall attack effectiveness. Future work may explore heuristic-guided or quality-aware initialization strategies to improve search performance. \textbf{Second}, the success of the attack also depends on the quality of the predefined strategy set. If the strategy pool contains a large number of low-quality or redundant attack actions, it may reduce the likelihood of identifying effective paths. Therefore, developing automatic filtering mechanisms and enhancing the overall quality of the strategy space are promising directions for further improvement.

\section{Conclusion}

In this study, we propose a dynamically weighted graph topology to characterize the multi-turn attack space, formalizing the dynamic search process of jailbreak attacks as a path planning problem. Building on this formulation, we develop ABC—an enhanced Artificial Bee Colony algorithm tailored for multi-turn attacks. ABC demonstrates significant advantages in both optimal path search and query cost control, enabling high attack success rates with substantially reduced query overhead. Extensive experimental results show that ABC achieves over 90\% ASR across five target models, requiring only 26 queries on average per successful attack, significantly outperforming existing baselines. Our work aims to facilitate the development of more robust defense mechanisms by providing an efficient and effective attack strategy, thereby offering a more comprehensive safeguard against jailbreak threats in real-world scenarios.


\bibliography{aaai2026}

\end{document}